\documentclass[11pt,a4paper]{article}
\usepackage[margin=1in]{geometry} 
\usepackage{color}
\usepackage{amsmath,titlesec,amsthm,amssymb,amsfonts,graphicx,siunitx,pifont,bbold,physics,slashed,empheq,bm,simplewick,multicol,enumitem}
\usepackage{jheppub}

\begin{document}

\title{\boldmath Generalized Carter \& R\"udiger Constants of $\sqrt{\text{Kerr}}$}

\author[1]{Chris de Firmian}

\affiliation[1]{Mani L. Bhaumik Institute for Theoretical Physics, University of California Los Angeles,\\ 430 Portola Plaza, Los Angeles, CA 90095, USA}

\emailAdd{cdefirmi@ucla.edu}

\author[2]{and Justin Vines}

\emailAdd{jvines@uark.edu}

\affiliation[2]{Department of Physics, University of Arkansas, 226 Physics
Building, 825 West Dickson Street, Fayetteville, AR 72701, USA}

\abstract{We consider the motion of a charged spinning test/probe particle --- governed by the Mathisson-Papapetrou-Dixon equations with generic, adiabatic, and conservative spin- and field-induced multipole moments --- in a background $\sqrt{\text{Kerr}}$ field on flat spacetime: the electromagnetic field of a charged spinning ring-disk singularity obtained from the $G\to 0$ limit of the Kerr-Newman solution for a charged spinning black hole. We investigate the existence of two extra \emph{hidden} constants of motion, analogous to the Carter constant (for geodesic motion in a Kerr spacetime, or for its spinning-probe generalization) and R\"udiger's linear-in-spin constant for a spinning probe in a Kerr background. We find that these two constants exist only when the Wilson coefficients parameterizing the probe's multipole structure take the particular values corresponding to ``spin-exponentiation'' of the effective Compton amplitudes through second order in spin.}

\maketitle
\flushbottom

\section{Introduction}

With the groundbreaking detection of gravitational waves marking a new era of observational astrophysics~\cite{LIGOScientific:2016aoc,LIGOScientific:2017vwq}, high precision measurements of compact astrophysical objects, such as black holes and neutron stars, have compelled significant theoretical progress in our understanding of classical two-body dynamics. In an effort to extend predictions beyond geodesic motion, amplitude- and worldline-based approaches have pushed the envelope in modeling real-world spin effects, which will play an increasingly important role as gravitational wave detectors become more sensitive \cite{Punturo:2010zz,  LISA:2017pwj,  Reitze:2019iox}. The study of generic spinning bodies in general relativity traces back to the 1950s \cite{Mathisson:1937zz,   Papapetrou:1951pa,   Pirani:1956tn,   Tulczyjew, Dixon:1970mpd, dixon2, dixon3}, with many successful field theoretic and worldline based approaches having since been developed to study spinning bodies in both the post-Newtonian (PN) approximation~\cite{Barker:1970zr, Barker:1975ae, Kidder:1992fr, Kidder:1995zr, Blanchet:1998vx, Tagoshi:2000zg, Porto:2005ac, Faye:2006gx, Blanchet:2006gy, Damour:2007nc, Steinhoff:2007mb, Levi:2008nh, Steinhoff:2008zr, Steinhoff:2008ji, Marsat:2012fn,Hergt:2010pa, Porto:2010tr,  Levi:2010zu,  Porto:2010zg, Levi:2011eq, Porto:2012as,  Hergt:2012zx,  Bohe:2012mr,  Hartung:2013dza,  Marsat:2013wwa,  Levi:2014gsa, Vaidya:2014kza, Bohe:2015ana,  Bini:2017pee,  Siemonsen:2017yux, Porto:2006bt,  Porto:2007tt,  Porto:2008tb,  Porto:2008jj,  Levi:2014sba, Levi:2015msa, Levi:2015uxa,  Levi:2015ixa,  Levi:2016ofk,   Levi:2019kgk,  Levi:2020lfn, Levi:2020kvb,  Levi:2020uwu,  Kim:2021rfj,  Maia:2017gxn,  Maia:2017yok,  Cho:2021mqw,  Cho:2022syn, Kim:2022pou,  Mandal:2022nty, Kim:2022bwv, Mandal:2022ufb, Levi:2022dqm,  Levi:2022rrq} and the post-Minkowskian (PM) approximation ~\cite{Bini:2017xzy, Bini:2018ywr, Maybee:2019jus,   Guevara:2019fsj, Chung:2020rrz,   Guevara:2017csg,   Vines:2018gqi,  Damgaard:2019lfh, Aoude:2020onz, Vines:2017hyw, Guevara:2018wpp, Chung:2018kqs, Chung:2019duq, Bern:2020buy, Kosmopoulos:2021zoq, Liu:2021zxr, Aoude:2021oqj, Jakobsen:2021lvp, Jakobsen:2021zvh, Chen:2021kxt, Chen:2022clh, Cristofoli:2021jas, Chiodaroli:2021eug, Cangemi:2022abk, Cangemi:2022bew, Haddad:2021znf, Aoude:2022trd, Menezes:2022tcs, Bern:2022kto, Chen:2022clh, Alessio:2022kwv, Bjerrum-Bohr:2023jau, Damgaard:2022jem, Haddad:2023ylx, Aoude:2023vdk, Jakobsen:2023ndj, Jakobsen:2023hig, Heissenberg:2023uvo, Bianchi:2023lrg, Bern:2023ity, gmoocv:2021, Ben-Shahar:2023djm}.

To model binary systems with spin, we employ the worldline formalism~\cite{Dixon:1970mpd,dixon2,dixon3} to describe one body as a probe or \emph{test body} in the background field of the other. Without spin, the probe's motion would be geodesic, however we can introduce deviation therefrom by modifying the geodesic action to include spin degrees of freedom. In the action, translational and rotational degrees of freedom, collectively referred to as \textit{minimal}, can be coupled to the background field in order to construct generic higher dimensional operators. As when working in effective field theories, one hierarchically organizes such operators, whether in powers of spin, coupling, or other perturbative parameters, to select only the \textit{relevant} interactions. By weighing each operator in the action by Wilson coefficients, one parameterizes families of probes, and the variation thereof yields effective equations of motion of a generic, minimal probe.

The resulting Mathisson-Papapetrou-Dixon (MPD) equations of motion~\cite{Mathisson:1937zz, Papapetrou:1951pa, Tulczyjew, Dixon:1970mpd, dixon2, dixon3}, governing the worldline and spin evolution, are said to be Liouville integrable if a sufficient number of conserved quantities exist. For a \textit{spinless}, gravitating probe in a Kerr spacetime background, the geodesic equation, $p^{\nu}\nabla_{\nu}p^{\mu}=0$, requires 4 independent conserved quantities to integrate on an 8-dimensional phase space. The Kerr spacetime background has 2 independent Killing vectors, associated with time-translation and azimuthal-rotation invariance, from which we have the conservation of energy and one component of angular momentum. Another constraint comes from worldline re-parameterization invariance in the form of the ``on shell'' condition $-m^2=g_{\mu\nu}p^{\mu}p^{\nu}$. Lastly, and unexpectedly, Carter~\cite{Carter:1968ab} found an additional, independently conserved constant $K_{\mu\nu}p^{\mu}p^{\nu}$, where $K_{\mu\nu}=K_{\nu\mu}$ is Killing tensor satisfying $\nabla_{(\mu}K_{\nu)\rho}=0$ --- a generalization of the Killing vector condition $\nabla_{(\mu}\xi_{\nu)}=0$. Thus geodesic motion in a Kerr spacetime is integrable, enabling the separation and reduction of the problem to one dimensional integrals.

Going beyond geodesic motion, we can generalize to a body with spin tensor $S_{\mu\nu}=-S_{\nu\mu}$, where we now have an additional 6 degrees of freedom. Via a Lagrange multiplier in the worldline action, one can enforce 3 additional constraints, $S_{\mu\nu}p^{\nu}=0$, called the spin supplementary condition (SSC), which can be understood as fixing the worldline along the object's center of mass. Additionally, the quadratic Casimir $\frac{1}{2}S_{\mu\nu}S^{\mu\nu}$ is independently conserved, leaving 2 unconstrained spin degrees of freedom. With suitable modifications, the on-shell constraint and two Killing vectors yield 3 additional conserved quantities. Lastly, for a black-hole-like probe, generalizations of Carter's constant along with another constant found by R\"udiger~\cite{Rudiger:1981,Rudiger:1983} were found to $\mathcal{O}(S^2)$~\cite{Comp_re_2023}. Consequently, with 8+2 degrees of freedom and 5 conserved quantities, the motion of a test black hole in a Kerr spacetime is integrable up to second order in spin. The focus of this paper will be to show an analogous result in an electromagnetic analog.

Complementary to the worldline formalism are amplitudes-based approaches to black hole scattering. Due to the computational complexity of gravity amplitudes, the amplitudes double copy is an indispensable tool~\cite{kosower2022sagexreviewscatteringamplitudes}. With its origins tracing back to Kawai, Lewellen \& Tye’s discovery of relations between open and closed string amplitudes~\cite{Kawai:1985xq}, and more recent color-kinematical formulations~\cite{Bern:2008qj}, the amplitudes double copy provides a prescription for constructing amplitudes of complicated theories from those of simpler theories. The squaring relation between the partial 3-point amplitudes of gluons and gravitons underpins the schematic Gravity = $(\text{Gauge Theory})^2$ mantra. The decades of work on the amplitudes double copy has produced a web of QFTs whose amplitudes are related, implying deeper connections between theories than what only their Lagrangians suggest at face value~\cite{Adamo:2022dcm}. Having revealed vast connections between the asymptotic observables of QFTs, a natural question is to ask whether analogous relations extend to the classical field configurations in the bulk.

For a class of solutions to Einstein's equations, called Kerr-Schild solutions~\cite{Kerr:Schild:1965}, one can decompose their metric as
\begin{align}
    g_{\mu\nu}=\eta_{\mu\nu}+\varphi k_\mu k_\nu,
\end{align}
where $\eta_{\mu\nu}$ is the Minkowski metric, and $k_{\mu}$ is a vector field which is null with respect to both the Minkowski and the full metric: $g^{\mu\nu}k_{\mu}k_{\nu}=0=\eta^{\mu\nu}k_{\mu}k_{\nu}$. The Kerr spacetime is such an example, where its $k_\mu$ exhibits a classical double copy relationship with an electromagnetic analog~\cite{Monteiro_2014}. Considering the Kerr-Newman solution, a charged and spinning black hole, one can take the $G\to 0$ limit, at fixed charge and fixed spin per mass, to recover a flat spacetime metric along with a nontrivial electromagnetic field, that of a charged spinning ring-disk singularity. This gauge field is of the form $A_{\mu}= \phi k_\mu$, where $\phi$ is a constant scalar multiple of $\varphi$, and $k_\mu$ is the same vector field as in the Kerr-Schild form of the Kerr metric. This electromagnetic gauge field, by virtue of this double-copy relation to Kerr, is called ``Root Kerr'' or $\sqrt{\text{Kerr}}$. Moreover, the $\sqrt{\text{Kerr}}$ field is also related to the Coulomb solution by the Newman-Janis shift~\cite{,Lynden-Bell:2002dvr}, a complex coordinate translation, analogous to how the Kerr solution derives from the Newman-Janis shift of the Schwarzschild solution~\cite{Newman:1965kpm}.

\begin{figure}[h]
  \centering
  \includegraphics[width=0.7\textwidth]{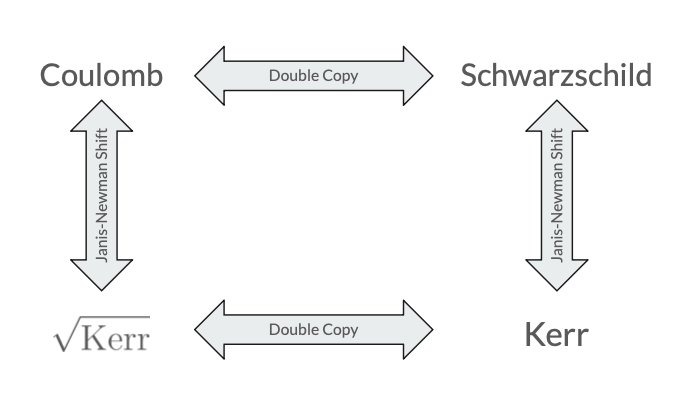}
  \caption{Double Copy \& Newman Janis Shift}
  \label{DoubleCopyJanisNewmanShift}
\end{figure}

The \emph{single copy} $\sqrt{\text{Kerr}}$ gauge field serves as an electromagnetic toy model for the gravitational double copy of Kerr, and, as in the gravitational case, we can treat a test $\sqrt{\text{Kerr}}$ probe in a $\sqrt{\text{Kerr}}$ background using the worldline formalism. Matching to the stationary multipole expansion of $\sqrt{\text{Kerr}}$ fixes a subset of the probe's Wilson coefficients \cite{Levi:2015msa}. The stationary field, however, does not fix so-called \emph{dynamical} multipole moments --- worldline operators characterizing the $\sqrt{\text{Kerr}}$ field's response to an external field. These dynamical multipole moments, perhaps to be computed through the Kerr-Newman solution's response to a linearized electromagnetic perturbation, are currently beyond the state of the art of black hole perturbation theory, such as Teukolsky-based approaches \cite{Bautista:2022wjf}. While currently undetermined, these multipole moments are nontheless important for understanding the $\sqrt{\text{Kerr}}$ dynamics, as in \cite{Scheopner:2024dim} where Scheopner \& Vines computed the electromagnetic Compton amplitude from a worldline action with generic coefficients. With the difficulty of fixing the dynamical moments, other proposed approaches, such as spin-shift symmetry~\cite{Aoude:2022trd, Bern:2022kto, Haddad:2023ylx,  Aoude:2023vdk}, higher-spin gauge symmetry \cite{Cangemi:2022bew}, and twistor worldline methods~\cite{Kim_2024}, have been suggested as principles for selecting a physically correct Compton amplitude. In this paper, we find generalized Carter \& R\"udiger constants to $\mathcal{O}(S^2)$, and show that their conservation is \textit{only} compatible with the stationary moments of $\sqrt{\text{Kerr}}$, up to $\mathcal{O}(S^2)$. Carter \& R\"udiger conservation also fixes the $\mathcal{O}(S^2)$ dynamical moments to zero, through which the $\sqrt{\text{Kerr}}$ Compton amplitude is constrained. The resulting Wilson coefficients continue the conjectured spin-exponentiation pattern~\cite{Guevara:2018wpp} of the Compton helicity amplitudes. 

\section{The Worldline Action \& the Electro-Magnetic MPD Equations}

\noindent To describe a charged, extended body, with translational, $z(\tau)$, and rotational, $e_{\hat{\mu}}^{\;\;\mu}(\tau)$, degrees of freedom, coupled to electromagnetism, $A_\mu(x)$, in flat spacetime, we can employ the following action \cite{dixon3, Ehlers:1977rud, bailey1975, Vines:2017hyw, Vines:2016unv, JanSteinhoff:2015ist, Marsat:2014xea, Levi:2015msa}:
\begin{align}
    \mathcal{S}[z,p,e,S,A]&=\int d\tau \left[(p_\mu+qA_\mu)\dot{z}^\mu+\frac{1}{2}S_{\mu\nu}\Omega^{\mu\nu}-\frac{\alpha}{2}(p^2+\mathcal{M}^2)-\beta^\mu S_{\mu\nu}p^\nu\right]\nonumber\\
    &-\frac{1}{16\pi}\int d^4x F_{\mu\nu}F^{\mu\nu}.
\end{align}
Here $\Omega^{\mu\nu}=e^{\hat{\mu}\mu}\dot{e}_{\hat{\mu}}^{\;\;\nu}$ is the angular velocity tensor, with conjugate spin tensor $S_{\mu\nu}$, and $p_{\mu}+qA_{\mu}$ conjugate to $\dot{z}^{\mu}$. The electromagnetic field strength tensor is the usual $F_{\mu\nu}=2\partial_{[\mu}A_{\nu]}$, and
\begin{align}
    \mathcal{M}^2&=\mathcal{M}^2(p,S,z)[F]=\mathcal{M}^2\left(p_\mu,S_{\mu\nu},\{F_{\mu\nu,N}(z)\}_{N=0}^\infty\right)
\end{align}
is the so-called \emph{dynamical mass function} squared, where $F_{\mu\nu,N}(z)=F_{\mu\nu,\rho_1...\rho_N}=\partial_{\rho_1}...\partial_{\rho_N}F_{\mu\nu}$. $\alpha$ and $\beta^\mu$ serve as Lagrange multipliers to enforce the ``on shell'' condition $p^2=-\mathcal{M}^2$ and the \emph{spin supplementary condition} $S_{\mu\nu}p^{\nu}=0$, respectively. The total variation, neglecting boundary terms, is
\begin{align}
    \delta \mathcal{S}&=\int d\tau\left\{\left[-\frac{d}{d\tau}(p_{\mu}+qA_{\mu})+q\dot{z}^{\nu}\partial_{\mu}A_{\nu}-\frac{\alpha}{2}\frac{\partial \mathcal{M}^2}{\partial z^{\mu}
    }\right]\delta z^{\mu}+\left[\dot{z}^{\mu}-\alpha p^{\mu}-\frac{\alpha}{2}\frac{\partial \mathcal{M}^2}{\partial p_{\mu}}-\beta^{\nu}S_{\nu}{}^{\mu}\right]\delta p_{\mu}\right.\nonumber\\
    &\quad\quad\quad\quad+\left.\frac{1}{2}\left[-\dot{S}_{\mu\nu}+2S_{\sigma[\mu}\Omega_{\nu]}{}^{\sigma}\right]e^{\hat{\rho}[\mu}\delta e_{\hat{\rho}}{}^{\nu]}+\frac{1}{2}\left[\Omega^{\mu\nu}-\alpha\frac{\partial \mathcal{M}^2}{\partial S_{\mu\nu}}-2\beta^{[\mu}p^{\nu]}\right]\delta S_{\mu\nu}\right\}\nonumber\\
    &\quad+\int d^4x\left[\frac{1}{4\pi}\partial_{\nu}F^{\nu\mu}+\int d\tau \left(q\dot{z}^{\mu}\delta^4(x-z)-\frac{\alpha}{2}\frac{\delta \mathcal{M}^2}{\delta A_{\mu}(x)}\right)\right]\delta A_{\mu},
\end{align}
where the variation of the frame is orthogonal: $\delta e_{\hat{\mu}}^{\;\;\mu} = e_{\hat{\mu}}^{\;\;\alpha} \delta\theta_\alpha^{\;\;\mu}$, with $\delta\theta^{\mu\nu} = -\delta\theta^{\nu\mu}$. We can further eliminate reference to $\Omega^{\mu\nu}$, $e^{\mu}{}_{\hat{\mu}}$, and $\beta^{\mu}$, along with solving algebraically for 
\begin{align}
    \alpha=p\cdot \dot{z}\left(p^2+\frac{p^{\mu}}{2}\frac{\partial \mathcal{M}^2}{\partial p^{\mu}}\right)^{-1},
\end{align}
such that the transport equations for $p_{\mu}$ and $S_{\mu\nu}$ become

\begin{align}
\dot{p}_\mu-q F_{\mu \nu} z^\nu & =\mathcal{F}_\mu=-\frac{\alpha}{2}\frac{\partial}{\partial z^{\mu}}\mathcal{M}^2\\
\dot{S}_{\mu \nu}-2 p_{[\mu} \dot{z}_{\nu]} & =\mathcal{N}_{\mu \nu}=-\alpha\left(p_{[\mu} \frac{\partial}{\partial p^{\nu]}}+2 S_{[\mu}{ }^\rho \frac{\partial}{\partial S^{\nu] \rho}}\right) \mathcal{M}^2,
\end{align}
subject to the SSC $p_{\mu}S^{\mu\nu}=0$.
We can further perform a change of variables by defining the momentum direction $u^\mu$ and the spin vector $s^\mu$ as
\begin{align}
    u^\mu= \frac{p^\mu}{\sqrt{-p^2}}, &\quad s^\mu=-\frac{1}{2}\epsilon^{\mu}_{\;\;\nu\rho\sigma}u^\nu S^{\rho\sigma},
\end{align}
which in turn implies
\begin{align}
    S^{\mu\nu}&=\epsilon^{\mu\nu}_{\;\;\;\;\rho\sigma}u^\rho s^\sigma.
\end{align}
Using our new variables, we take $\mathcal{M}^2(p,S,z)\to \mathcal{M}^2(u,s,z)$, i.e., we trade the spin tensor $S$ for the spin vector $s$:
\begin{align}
    \left(\frac{\partial}{\partial p_\mu}\right)_S\mathcal{M}^2(p,s(p,S))&=\left[\left(\frac{\partial}{\partial p_\mu}\right)_s+\left(\frac{\partial s^\nu}{\partial p_\mu}\right)_S\left(\frac{\partial}{\partial s^\nu}\right)_p\right]\mathcal{M}^2(p,s),\\
    \left(\frac{\partial}{\partial S_{\mu\nu}}\right)_p\mathcal{M}^2(p,s(p,S))&=\left(\frac{\partial s^\rho}{S_{\mu\nu}}\right)_p\left(\frac{\partial }{\partial s^\rho}\right)_p\mathcal{M}^2(p,s).
\end{align}
Moreover, defining the orthogonal projector to $u$, $\mathcal{P}$,
\begin{align}
    \eta^{\mu\nu}&=\frac{\partial p^\mu}{\partial p_\nu}=\frac{\partial }{\partial p_\nu}\left(\mathcal{M}u^\mu\right)=\mathcal{M}\frac{\partial u^\mu}{\partial p_\nu}+u^\mu\frac{\partial \mathcal{M}}{\partial p_\nu}\\
    \Longrightarrow \mathcal{M}\frac{\partial u^\mu}{\partial p_\nu}&= \eta^{\mu\nu}-u^\mu\frac{\partial \sqrt{-p^2}}{\partial p_\nu}=\eta^{\mu\nu}-u^\mu \frac{-1}{2\mathcal{M}}\frac{\partial p^2}{\partial p_\nu}=\eta^{\mu\nu}+u^\mu u^\nu=:\mathcal{P}^{\mu\nu},
\end{align}
we can take $\mathcal{M}^2$ to depend on $p$ \emph{only through its direction} $u$,
\begin{align}
    \frac{\partial \mathcal{M}^2}{\partial p^\mu}&=\frac{\partial u^\nu}{\partial p^\mu}\frac{\partial \mathcal{M}^2}{\partial u^\nu}=\mathcal{P}^{\nu}_{\;\;\mu}\frac{1}{\mathcal{M}}\frac{\partial \mathcal{M}^2}{\partial u^\nu}=\mathcal{P}^\nu_{\;\;\mu}2\frac{\partial \mathcal{M}}{\partial u^\nu}
\end{align}
in which case
\begin{align}
    p^\mu\frac{\partial \mathcal{M}^2}{\partial p^\mu}=p^\mu \mathcal{P}^\nu_{\;\;\mu}2\frac{\partial \mathcal{M}}{\partial u^\nu}=\left(p^\nu-p^\nu\right)2\frac{\partial \mathcal{M}}{\partial u^\nu}&=0\Longrightarrow \alpha=\frac{-p^\mu \dot{z}_\mu}{-p^2}=-\frac{u^\mu \dot{z}_\mu}{\mathcal{M}}.
\end{align}
We can fix the worldline parameterization using Dixon's condition, $u\cdot\dot{z}=-1$ such that $\alpha=1/\mathcal{M}$. The resulting equations of motion are
\begin{align}
\dot{p}_\mu-q F_{\mu \nu} \dot{z}^\nu &=\mathcal{F}_\mu =-\frac{\alpha}{2} \frac{\partial}{\partial z^\mu} \mathcal{M}^2 ,
\\
\dot{S}_{\mu \nu}-2 p_{[\mu} \dot{z}_{\nu]} &=\mathcal{N}_{\mu \nu}=-\alpha\left(p_{[\mu} \frac{\partial}{\partial p^{\nu]}}+s_{[\mu} \frac{\partial}{\partial s^{\nu]}}\right) \mathcal{M}^2,
\\
\dot{s}^{\mu}&=
\frac{\alpha}{2}\left(\epsilon^{\mu \nu}{ }_{\rho \sigma} u^\rho s^\sigma \frac{\partial}{\partial s^\nu}-\frac{u^\mu}{\mathcal{M}} s^\nu \frac{\partial}{\partial z^\nu}\right) \mathcal{M}^2+\frac{u^\mu}{\mathcal{M}} q F_{\nu \rho} s^\nu \dot{z}^\rho,
\\
\dot{z}^{\mu}&=u^{\mu}-\frac{\mathcal{N}^{\mu\nu}u_{\nu}}{\mathcal{M}}-S^{\mu\nu}\frac{\mathcal{F}_{\nu}+qF_{\nu\rho}(u^{\rho}-\mathcal{N}^{\rho\sigma}u_{\sigma}/\mathcal{M})}{\mathcal{M}^2-\frac{1}{2}qF_{\alpha\beta}S^{\alpha\beta}}.
\end{align}

\section{The \texorpdfstring{$\sqrt{\text{Kerr}}$}{Root Kerr} Field}

The $\sqrt{\text{Kerr}}$ gauge field can be derived from the Kerr-Newman solution. Consider the Kerr-Newman metric and vector potential in Boyer-Lindquist coordinates $(t,r,\theta,\phi)$,
\begin{align}
g_{\mu \nu} \mathrm{d} x^\mu \mathrm{d} x^\nu & =-\frac{\Delta}{\Sigma}\left(\mathrm{d} t-a \sin ^2 \theta \mathrm{~d} \phi\right)^2+\frac{\Sigma}{\Delta} \mathrm{d} r^2+\Sigma \mathrm{d} \theta^2+\frac{\sin ^2 \theta}{\Sigma}\left(\left(r^2+a^2\right) \mathrm{d} \phi-a \mathrm{~d} t\right)^2 \nonumber\\
A_\mu \mathrm{d} x^\mu & =\frac{Q r}{\Sigma}\left(\mathrm{d} t-a \sin ^2 \theta \,\mathrm{d} \phi\right),
\end{align}
with
\begin{align}
\Sigma=r^2+a^2 \cos ^2 \theta, \quad \Delta=r^2+a^2+G Q^2-2 G M r,
\end{align}
outside a black hole with mass $M$, spin $S=Ma$, and electric charge $-Q$. Taking the $G\to 0$ limit we have
\begin{align}
\eta_{\mu \nu} \mathrm{d} x^\mu \mathrm{d} x^\nu & =-\mathrm{d} t^2+\frac{\Sigma}{r^2+a^2} \mathrm{~d} r^2+\Sigma \mathrm{d} \theta^2+\left(r^2+a^2\right) \sin ^2 \theta \mathrm{~d} \phi^2,
\\
A_\mu \mathrm{d} x^\mu & =\frac{Q r}{\Sigma}\left(\mathrm{d} t-a \sin ^2 \theta \,\mathrm{d} \phi\right),
\end{align}
namely a flat metric in oblate-spheroidal coordinates along with the $\sqrt{\text{Kerr}}$ gauge field. We can relate these to the Cartesian coordinates as
\begin{align}
t=t,\quad x+i y=\sqrt{r^2+a^2} \sin \theta\, e^{i\phi}, \quad z=r \cos \theta,
\end{align}
implying
\begin{align}
(r+i a \cos \theta)^2=x^2+y^2+(z+i a)^2=|\boldsymbol{x}+i \boldsymbol{a}|^2.
\end{align}
Defining, via Cartesian components, the time-like Killing vector $T^{\mu}:=(1,0,0,0)$, the (rescaled) spin vector $a^{\mu}:=(0,0,0,a)$, and the spacetime displacement from the origin $x^{\mu}:=(t,x,y,z)$, then $|\mathbf{x}|:= \sqrt{x_{\mu}x^{\mu}+(T_{\mu}x^{\mu})^2}$ is the magnitude of the projection of $x^\mu$ orthogonal to $T^\mu$. With this and
\begin{align}
\mathcal{R}:=r+i a \cos \theta=|\boldsymbol{x}+i \boldsymbol{a}|=\sqrt{(x+i a)_\mu(x+i a)^\mu+\left[T_\mu(x+i a)^\mu\right]^2},
\end{align}
it can be shown that
\begin{align}
A_\mu&=\left[T_\mu \cos (a \cdot \partial)+\epsilon_{\mu \nu \alpha \beta} T^\nu a^\alpha \partial^\beta \frac{\sin (a \cdot \partial)}{a \cdot \partial}\right] \frac{-Q}{|\boldsymbol{x}|},
\\
F_{\mu\nu}&=2\partial_{[\mu}A_{\nu]}=\frac{1}{2} G_{\mu \nu}{ }^{\alpha \beta} T_\alpha(x+i a)_\beta \frac{-Q}{\mathcal{R}^3}+c . c .,
\end{align}
where $G_{\mu \nu}{ }^{\alpha \beta}:=2 \delta_{[\mu}{ }^\alpha \delta_{\nu]}{ }^\beta+i \epsilon_{\mu \nu}{ }^{\alpha \beta}$ is the tensor which projects a 2-form onto 4 times its anti-self dual part. An important building block for working with the $\sqrt{\text{Kerr}}$ background the anti-self dual 2-form 
\begin{align}
    N_{\mu \nu}:=i G_{\mu \nu}{ }^{\alpha \beta} T_\alpha \partial_\beta \mathcal{R}=i G_{\mu \nu}{ }^{\alpha \beta} T_\alpha(x+i a)_\beta \frac{1}{\mathcal{R}},
\end{align}
as it allows us to re-write $F_{\mu\nu}$ as
\begin{align}
    F_{\mu \nu}=\frac{i Q}{2 \mathcal{R}^2} N_{\mu \nu}-\frac{i Q}{2 \overline{\mathcal{R}}^2} \bar{N}_{\mu \nu},
\end{align}
along with expressing the Killing-Yano tensor
\begin{align}
    Y_{\mu \nu}=\frac{1}{2}\left(\mathcal{R} N_{\mu \nu}+\overline{\mathcal{R}} \bar{N}_{\mu \nu}\right),
\end{align}
from which the Killing tensor $K_{\mu\nu}=Y_{\mu\rho}Y^{\rho}{}_{\nu}$ can be built.
$N$ and its complex conjugate $\bar{N}$ can be thought of as ``square roots of the metric'' in the sense
\begin{align}
    N_{\mu \rho} N_\nu{ }^\rho=\eta_{\mu \nu}=\bar{N}_{\mu \rho} \bar{N}_\nu{ }^\rho.
\end{align}
Moreover the tensor $\pi_{\mu\nu}:=N_{\mu\rho}\bar{N}^{\rho}{}_{\nu}$ satisfies $\pi_{\mu\nu}=\pi_{\nu\mu}=\bar{\pi}_{\mu\nu}$ and $\pi_{\mu}{}^{\mu}=0$. 

\noindent Lastly, these building blocks have the following closed set of differential relations:
\begin{align}
    \partial_\alpha N_{\mu \nu}&=\frac{-i}{\mathcal{R}}\left(G_{\mu \nu \alpha \beta}-N_{\mu \nu} N_{\alpha \beta}\right) T^\beta,
    \\
    \partial_\mu \mathcal{R}&=-i N_{\mu \nu} T^\nu,
    \\
    \partial_\alpha Y_{\mu \nu}&=\epsilon_{\mu \nu \alpha \beta} T^\beta,
\end{align}
given $\partial_\alpha G_{\mu\nu\rho\sigma}=0$.

\section{Carter \& R\"udiger Ans\"atze}

\subsection{\texorpdfstring{$\mathcal{O}(S^2)$}{OS2} Ansatz}

To write an ansatz for the Carter and R\"udiger constants, $C$ and $Q$, we start with the two guiding principles: parity and scaling dimension. Requiring that both constants be scalars, not \emph{pseudo}-scalars, restricts each term in the ansatz to be parity-even. As for their scaling dimensions, $C$ and $Q$ have dimensions $[M]^2[L]^2$ and $[M][L]^2$, respectively. The ans\"atze, however, can be combined into one by rescaling terms with the probe mass $m$. To minimize overcounting, we introduce the electric and magnetic fields in the $u^\mu$ frame,
\begin{align}
    E_\mu&=-u^{\nu}F_{\nu\mu},\quad B_\mu=u^{\nu}{}^*F_{\nu\mu},
\end{align}
along with the derivatives in the $u^\mu$ direction,
\begin{align}
    \dot{E}_\mu &= -u^{\nu}u^{\rho}\partial_\rho F_{\nu\mu},\quad \dot{B}_\mu = u^{\nu}u^{\rho}\partial_{\rho}{}^*F{}_{\nu\mu},
\end{align}
with ${}^*$ denoting the Hodge dual. Moreover, we also introduce the symmetric tensors
\begin{align}
    \mathcal{E}_{\mu\nu}&=-u^{\sigma}\partial_{\rho}F_{\sigma(\mu}\mathcal{P}_{\nu)}{}^{\rho},\quad \mathcal{B}_{\mu\nu}=u^{\sigma}\partial_{\rho}{}^*F_{\sigma(\mu}\mathcal{P}_{\nu)}{}^{\rho},
\end{align}
where $\mathcal{P}^{\mu\nu}=\eta^{\mu\nu}+u^{\mu}u^{\nu}$ is the perpendicular projector onto $u$, along with the vectors
\begin{align}
    X_{\mu}&=-u^{\nu}{}^*Y_{\nu\mu},\quad W_{\mu}=-u^{\nu}Y_{\nu\mu}.
\end{align}
Organized in terms of powers of $s$, $q$ and $T$, we took our ansatz to be a real linear combination of the following terms:

\begin{itemize}[leftmargin=*]
  \item[$\mathcal{O}(s^0)$\;\;\quad]\begin{itemize}[]
    \item[$\mathcal{O}(q^0)$\quad]\begin{itemize}[]
        \item[$\mathcal{O}(T^0)$]\quad $p^2W^2\oplus p^2 X^2$
        \end{itemize}
    \end{itemize}
  \item[$\mathcal{O}(s^1)$\;\;\quad]\begin{itemize}
    \item[$\mathcal{O}(q^0)$\quad]\begin{itemize}
        \item[$\mathcal{O}(T^0)$]\quad $m\;s\cdot W$
        \item[$\mathcal{O}(T^1)$]\quad $p\cdot T \;s\cdot W\oplus \epsilon[TXps]$
    \end{itemize}
    \item[$\mathcal{O}(q^1)$\quad]\begin{itemize}
        \item[$\mathcal{O}(T^0)$]\quad \begin{equation}
            q\left(\begin{matrix}
        X^2 \;B\cdot s\\
        X\cdot B\; X\cdot s\\
        W^2 \;B\cdot s\\
        W\cdot B \;W\cdot s
        \end{matrix}\quad\oplus\quad\begin{matrix}
            X\cdot W\; E\cdot s\\
            W\cdot E \;X\cdot s\\
            X\cdot E \;W\cdot s
        \end{matrix}\quad\oplus\quad \epsilon[XWBs]\right)\nonumber
        \end{equation}
        \end{itemize}
    \end{itemize}
  \item[$\mathcal{O}(s^2)$\;\;\quad]\begin{itemize}
      \item[$\mathcal{O}(q^0)$\quad]\begin{itemize}
          \item[$\mathcal{O}(T^1)$]\quad $s^2 \;T\cdot u$
          \item[$\mathcal{O}(T^2)$]\quad $(s\cdot T)^2\oplus s^2 (T\cdot u)^2$
      \end{itemize}
      \item[$\mathcal{O}(q^1)$\quad]\begin{itemize}
          \item[$\mathcal{O}(T^0)$]\quad \begin{equation}\dfrac{q}{m}\left(\begin{matrix}
              s^2\; E\cdot X\\
              E\cdot s\; s\cdot X\\
              s^2\; B\cdot W\\
              B\cdot s\; s\cdot W
          \end{matrix}\quad\oplus\quad\begin{matrix}
              X\cdot\mathcal{E}\cdot X\; s^2\\
              s\cdot\mathcal{E}\cdot s \;X^2\\
              X\cdot \mathcal{E}\cdot s\; X\cdot s\\
              W\cdot\mathcal{E}\cdot W\; s^2\\
              s\cdot\mathcal{E}\cdot s \;W^2\\
              W\cdot \mathcal{E}\cdot s\; W\cdot s
          \end{matrix}\quad\oplus\quad\begin{matrix}
              X\cdot \mathcal{B}\cdot W\; s^2\\
              X\cdot \mathcal{B}\cdot s\; W\cdot s\\
              W\cdot \mathcal{B}\cdot s\; X\cdot s\\
              s\cdot \mathcal{B}\cdot s\; X\cdot W
          \end{matrix}\quad\oplus\quad\begin{matrix}
              \epsilon[u\dot{E}XW]\;s^2\\
              \epsilon[su\dot{E}X]\; W\cdot s\\
              \epsilon[Wsu\dot{E}]X\cdot s\\
              \epsilon[XWsu]\;\dot{E}\cdot s\\
              \epsilon[u\dot{B}Xs]\;X\cdot s\\
              \epsilon[u\dot{B}Ws]\; W\cdot s
          \end{matrix}\right)\nonumber\end{equation}
          \item[$\mathcal{O}(T^1)$]\quad \begin{equation}\dfrac{q}{m}\left(\begin{matrix}
              \epsilon[XBsu]\; T\cdot s\\
              \epsilon[BsuT]\; X\cdot s\\
              \epsilon[suTX]\; B\cdot s\\
              \epsilon[uTXB]\; s^2\\
          \end{matrix}\quad\oplus\quad\begin{matrix}
              \epsilon[WEsu]\; T\cdot s\\
              \epsilon[EsuT]\;W\cdot s\\
              \epsilon[suTW]\; E\cdot s\\
              \epsilon[uTWE]\; s^2\\
          \end{matrix}\quad\oplus\quad \left(T\cdot u\right)\left[\begin{matrix}
              X\cdot E\; s^2\\
              X\cdot s\; E\cdot s\\
              W\cdot B\; s^2\\
              W\cdot s\; B\cdot s
          \end{matrix}\right]\right)\nonumber\end{equation}
      \end{itemize}
      \item[$\mathcal{O}(q^2)$\quad]\begin{itemize}
          \item[$\mathcal{O}(T^0)$]\quad \begin{equation}\dfrac{q^2}{m^2}\left(\begin{matrix}
              E\cdot s\; E\cdot X\; X\cdot s\\
              E^2\; (X\cdot s)^2\\
              (E\cdot s)^2\; X^2\\
              s^2\;(E\cdot X)^2\\
              s^2\; E^2\; X^2\\
              E\cdot s\; E\cdot W\; W\cdot s\\
              E^2\; (W\cdot s)^2\\
              (E\cdot s)^2\; W^2\\
              s^2\;(E\cdot W)^2\\
              s^2\; E^2\; W^2
          \end{matrix}\quad\oplus\quad\begin{matrix}
              B\cdot s\; B\cdot X\; X\cdot s\\
              B^2\; (X\cdot s)^2\\
              (B\cdot s)^2\; X^2\\
              s^2\;(B\cdot X)^2\\
              s^2\; B^2\; X^2\\
              B\cdot s\; B\cdot W\; W\cdot s\\
              B^2\; (W\cdot s)^2\\
              (B\cdot s)^2\; W^2\\
              s^2\;(B\cdot W)^2\\
              s^2\; B^2\; W^2
          \end{matrix}\quad\oplus\quad\begin{matrix}
              s\cdot E\; E\cdot X\; X\cdot s\\
              s\cdot E\; B\cdot W\; X\cdot s\\
              s\cdot X\; B\cdot E\; W\cdot s\\
              s\cdot B\; E\cdot X\; W\cdot s\\
              s\cdot B\; E\cdot W\; X\cdot s\\
              s\cdot E\; X\cdot W\; B\cdot s\\
              s^2\; E\cdot B\; X\cdot W\\
              s^2\; E\cdot X\; B\cdot W\\
              s^2\; E\cdot W\; X\cdot B
          \end{matrix}\right)\nonumber\end{equation}
      \end{itemize}
  \end{itemize}
\end{itemize}

In the above we have used the shorthand $\cdot$ to denote Lorentz index contraction, squaring to denote 4-vector norm, and $\epsilon[wxyz]=\epsilon^{\mu\nu\rho\sigma}w_\mu x_\nu y_\rho z_\sigma$. Each term listed, whether included in a column or separated by a direct sum symbol $\oplus$, is included as part of the ansatz, with each term being weighed by a dimensionless coefficient.

\subsection{Reduction to Basis}

To solve for the Wilson coefficients that were compatible with $C$ and $Q$ conservation, we reduced the results of taking $\frac{dC}{d\tau}$ and $\frac{dQ}{d\tau}$ to a basis spanned by $u^{\mu}$, $N^{\mu\nu}u_{\nu}$, $\bar{N}^{\mu\nu}u_{\nu}$, and $\pi^{\mu\nu}u_{\nu}$, where $\Pi^{\mu\nu}=N^{\mu\rho}\bar{N}_{\rho}{}^{\nu}$. To do so, we replaced all scalars not involving a contraction of $u$ by the use of several Shouten-like identities,
\begin{align}
    s\cdot T&= \frac{N_{su}N_{Tu}+\bar{N}_{su}\bar{N}_{Tu}+\pi_{su}\pi_{Tu}+(T\cdot u) \pi_{su}\pi_{uu}-\bar{N}_{su}N_{Tu}\pi_{uu}-N_{su}\bar{N}_{Tu}\pi_{uu}}{-1+(\pi_{uu})^2}\\
    \pi_{sT}&=\frac{(T\cdot u)\pi_{su}-\bar{N}_{su}N_{Tu}-N_{su}\bar{N}_{Tu}+N_{su}N_{Tu}\pi_{uu}+\bar{N}_{su}\bar{N}_{Tu}\pi_{uu}+\pi_{su}\pi_{Tu}\pi_{uu}}{-1+(\pi_{uu})^2}\\
    N_{sT}&=\frac{(T\cdot u)N_{su}+\pi_{su}\bar{N}_{Tu}-\bar{N}_{su}\pi_{Tu}-(T\cdot u)\bar{N}_{su}\pi_{uu}-\pi_{su}N_{Tu}\pi_{uu}+N_{su}\pi_{Tu}\pi_{uu}}{-1+(\pi_{uu})^2}\\
    \bar{N}_{sT}&=\frac{(T\cdot u)\bar{N}_{su}+\pi_{su}N_{Tu}-N_{su}\pi_{Tu}-(T\cdot u)N_{su}\pi_{uu}-\pi_{su}\bar{N}_{Tu}\pi_{uu}+\bar{N}_{su}\pi_{Tu}\pi_{uu}}{-1+(\pi_{uu})^2}\\
    s\cdot s&= \frac{(N_{su})^2+(\bar{N}_{su})^2+(\pi_{su})^2-2N_{su}\bar{N}_{su}\pi_{uu}}{-1+(\pi_{uu})^2}\\
    \pi_{ss}&=\frac{-2N_{su}\bar{N}_{su}+(N_{su})^2\pi_{uu}+(\bar{N}_{su})^2\pi_{uu}+(\pi_{su})^2\pi_{uu}}{-1+(\pi_{uu})^2},
\end{align}
where we have used the shorthand notation to denote the vector contraction with a tensor as follows: $\pi_{ss}=\pi_{\mu\nu}s^{\mu}s^{\nu}$, $N_{su}=N_{\mu\nu}s^{\mu}u^{\nu}$, $\bar{N}_{Tu}=\bar{N}_{\mu\nu}T^{\mu}u^{\nu}$, et cetera. Lastly, the relation
\begin{align}
    (\pi_{uu})^2&=1-(T\cdot u)^2-(N_{Tu})^2-(\bar{N}_{Tu})^2-(\pi_{Tu})^2+2N_{Tu}\bar{N}_{Tu}\pi_{uu}-2(T\cdot u)\pi_{Tu}\pi_{uu}
\end{align}
proves useful for occasional simplification. Reducing the $\tau$-derivative of our ansatz to this basis and setting the individual terms' coefficients to zero allows us to solve for the coefficients of the ansatz and also to restrict the Wilson coefficients in the dynamical mass function.

\section{Carter, R\"udiger Results \& Implications for Compton Scattering}

In our action, the most general dynamical mass function to $\mathcal{O}(S^2)$ is
\begin{align}
    \mathcal{M}^2&=m^2+C_1 2q {}^*F_{\mu\nu}s^{\mu}u^{\nu}+C_2\frac{q}{m}s^{\mu}s^{\nu}u^{\rho}\nabla_{\mu}F_{\nu\rho}\\
    &+D_1\frac{q^2}{m^2}\left(F_{\mu\nu}u^\mu s^{\nu}\right)^2+D_2\frac{q^2}{m^2}\left({}^*F_{\mu\nu}s^{\mu}u^{\nu}\right)^2+D_3\frac{q^2}{m^2}s^{\mu}s_{\mu}F^{\rho\sigma}F_{\rho\sigma}+D_4\frac{q^2}{m^2}s^{\mu}s_{\mu}u^{\nu}F_{\nu\rho}F^{\rho\sigma}u_\sigma\nonumber,
\end{align}
where $m$ is the mass of the spin-less probe. In addition to the conserved quantities resulting from our two Killing vectors $\xi\in\{t,e_\phi\}$,
\begin{align}
    \mathcal{P}_{\xi}=\left(p_\mu+q A_\mu\right) \xi^\mu+\frac{1}{2} S^{\mu \nu} \partial_\mu \xi_\nu,
\end{align}
we find that, to $\mathcal{O}(S^2)$, the following generalizations of Carter and R\"udiger's constants are conserved:
\begin{align}
    C&=Y_{\mu \rho} Y_\nu{ }^\rho p^\mu p^\nu - q F^\mu{ }_\sigma S^{\nu \sigma}Y_{\mu \rho} Y_\nu{ }^\rho+4 T^\lambda \epsilon_{\lambda \rho \sigma[\mu} Y_{\nu]}{ }^\sigma p^\mu S^{\nu \rho}\nonumber\\
    &-(s^\mu T_{\mu})^2+s^\mu s_{\mu} (T^\nu u_{\nu})^2-\frac{q}{m}s^\mu s_{\mu} T^\nu u_\nu F^{\rho\sigma}{}^*Y_{\rho\sigma}-\frac{2q}{m}s^\mu T_\mu F^{\rho\sigma}s^\nu u_\rho {}^*Y_{\nu\sigma}\nonumber\\
    &+\frac{3q}{m}T^\mu u_\mu F_{\nu}{}^{\rho}s^{\nu}s^{\sigma}{}^*Y_{\sigma\rho}+\frac{q}{m}s^{\mu}s_{\mu}F_{\nu}{}^{\rho}T^{\nu}u^{\sigma}{}^*Y_{\sigma\rho}+\frac{q}{m}s^{\mu}s^{\nu}u^{\rho}{}^*Y_{\rho}{}^{\sigma}{}^*Y_{\sigma}{}^{\gamma}\nabla_{\nu}F_{\mu\gamma},
    \end{align}
and
\begin{align}
    Q&=\frac{1}{2}S^{\mu\nu}{}^*Y_{\mu\nu}+\frac{3q}{2m^2}F_{\mu\nu}s^{\mu}u^{\nu}s^{\rho}u^{\sigma}{}^*Y_{\rho\sigma}+\frac{q}{2m^2}F_{\mu\nu}s^{\mu}s^{\rho}{}^*Y_{\rho}{}^{\nu}-\frac{q}{m^2}s^{\mu}s_{\mu}F_{\alpha}{}^{\gamma}u^{\alpha}u^{\beta}{}^*Y_{\beta\gamma},
\end{align}
only for a unique choice of the Wilson coefficients in the action.

The conservation of $C$ and $Q$ fixes the multipole coefficients $C_1=C_2=1$, in agreement with the $\sqrt{\text{Kerr}}$ stationary multipole moments, but also fixes the $D_1=D_2=D_3=D_4=0$ dynamical multipole moments. This implies that the resulting Compton helicity amplitudes, as derived by Scheopner and Vines~\cite{Scheopner:2024dim}, exhibit the ``spin-exponentiation'' property up to second order.
Concretely, consider Compton scattering in a particular Lorentz frame, associated to inertial Cartesian coordinates $x^\mu=(t,x,y,z)$, where $\sqrt{\text{Kerr}}$ has initial velocity $v^\mu=(1,0,0,0)$ and the incoming and outgoing photons have wavevectors $k_1^\mu=\omega(1,0,0,1)$ and $k_2^\mu=\omega(1,\sin\theta,0,\cos\theta)$, respectively. In terms of the squared momentum transfer $q^2=(k_2-k_1)^2=4 \omega^2 \sin ^2 \frac{\theta}{2}$, the helicity preserving $A_{++}$ and reversing $A_{+-}$ Compton amplitudes~\cite{Scheopner:2024dim} are 
\begin{align}
    \mathcal{A}_{++}&=\frac{q^2-4\omega^2}{2\omega^2}e^{(k_1+k_2-2w)\cdot a}+\mathcal{O}(S^3)\\
    \mathcal{A}_{+-}&=\frac{q^2}{2\omega^2}e^{(k_1-k_2)\cdot a}+\mathcal{O}(S^3),
\end{align}
where $\omega=-v\cdot k_1=-v\cdot k_2$, and $w^\mu=\omega\left(1, \tan \frac{\theta}{2}, i \tan \frac{\theta}{2}, 1\right)$.
\section{Discussion, Conjectures \& Future Directions}

With our identification of Carter and R\"udiger's constants, a sufficient number of independently conserved quantities exist for the dynamics of a particular $\sqrt{\text{Kerr}}$ probe in a $\sqrt{\text{Kerr}}$ background to be integrable to $\mathcal{O}(S^2)$. The integrability of a $\sqrt{\text{Kerr}}$ probe to $\mathcal{O}(S^2)$ opens the door for the Dirac-Bracket formalism \cite{Gonzo:2024sbo} to compute impulse and spin kicks, as was done for Kerr in \cite{akpinar2025unexpectedsymmetrieskerrblack}. With integrability picking out, order by order in spin, the  Kerr~\cite{Comp_re_2023} and $\sqrt{\text{Kerr}}$ probes in Kerr and $\sqrt{\text{Kerr}}$ backgrounds, respectively, we suspect that the $\mathcal{O}(S^3)$ generalizations of the electromagnetic Carter \& R\"udiger constants, are they found to exist, will also affix the \emph{stationary} $\mathcal{O}(S^3)$ multipole moments to those of $\sqrt{\text{Kerr}}$.

As for the dynamical moments, we conjecture that Liouville integrability will continue to fix higher order multipole moments. These coefficients would pick particular Compton amplitudes, and while the helicity preserving amplitude $A_{++}$ is known to not exponentiate past $\mathcal{O}(S^2)$\cite{Guevara:2018wpp,Chung:2018kqs}, the helicity reversing Compton amplitude $A_{+-}$ encounters no such limitation and may yet exponentiate to $\mathcal{O}(S^3)$.

\acknowledgments

Many thanks to Zvi Bern for his support, insightful discussions and input on the draft of this paper. Many thanks to Trevor Scheopner for many enlightening conversations. We are also grateful to the Mani L. Bhaumik Institute for Theoretical Physics for support. This work was supported in part by the U.S. Department of Energy (DOE) under award number DE-SC0009937.

%%%
\bibliography{ref}{}
\setlength{\bibsep}{0pt plus 0.1ex}
\bibliographystyle{JHEP}
%%%

\end{document}